\documentstyle[12pt]{article}
\def\ie{\mbox{\it i.e.\ \rm}}
\def\eg{\mbox{\it e.g.\ \rm}}

\def\be{\begin{equation}}
\def\ee{\end{equation}}
\def\picture #1 by #2 (#3){
  \vbox to #2{
    \hrule width #1 height 0pt depth 0pt
    \vfill
    \special{picture #3} 
    }
  }
\def\scaledpicture #1 by #2 (#3 scaled #4){{
  \dimen0=#1 \dimen1=#2
  \divide\dimen0 by 1000 \multiply\dimen0 by #4
  \divide\dimen1 by 1000 \multiply\dimen1 by #4
  \picture \dimen0 by \dimen1 (#3 scaled #4)}
  }

\begin{document}
%
%
%
%
\begin{titlepage}

\title{\bf Residual long-range pseudoscalar forces between unpolarised
 macroscopic bodies\vspace{2cm}}

\author{
{\bf J. A.\ Grifols} \\ {\bf S.\ Tortosa }\vspace{.5cm}\\IFAE Institut de
F{\'{\i}}sica d'Altes Energies \\ Grup de F{\'{\i}}sica Te{\`o}rica\\
Universitat Aut{\`o}noma de
Barcelona\\ 08193 Bellaterra, Barcelona, Catalonia, Spain} \date{\null}
\maketitle

\begin{abstract}
In this paper we survey the effects of residual long-range forces
associated to $\gamma_5$-spin dependent-couplings of fermions to
massless bosons exerted by unpolarised bulk matter over macroscopic
distances. We establish that such forces with behaviour proportional to
$R^{-6}$ do indeed exist. They arise as a quantum mechanical effect
due to simultaneous exchange of two quanta. We explore their presence
in existing astronomical as well as laboratory data on non-newtonian
components of the force between macroscopic bodies. Since no limits on
their real existence could be found, we conclude that residual
long-range pseudoscalar attractive and composition dependent forces between
neutral
unpolarised bulk matter extending over macroscopic distances are very
efficiently
shielded over a huge range of distances: from astronomical scales down
to the micron scale.
 \end{abstract} \end{titlepage}

The issue whether long-range interactions other than the already known
in nature exist has been raised repeatedly and from different fronts of both
experimental/observational and theoretical physics. To name just a few, the
reanalysis of E{\"o}tv{\"o}s experiment and all the subsequent experimental and
theoretical activity boosted by a $5^{\mbox{\scriptsize \rm th}}$ force
hypothesis \cite{a}; the dark
matter problem and its feeble non gravitational interactions \cite{b}; the
hypothesised
existence of very light particles (\eg axions, majorons, etc.) to solve
profound  questions in particle physics, and which could mediate new
interactions
or the idea that baryon number (or
lepton number) might be charges associated to a local gauge symmetry
and thus long-range interactions should be associated to the exchange of
massless vector quanta \cite{c}.
A vast domain of couplings and ranges for such forces has been explored and
limits thereof have been set in terrestrial laboratories as well as
astronomical/astrophysical/cosmological environments \cite{d}. Mostly the
long-range
interactions for which astronomical or laboratory limits do exist are
associated to vector or scalar quanta. Pseudoscalar interactions, on the other
hand, cannot be directly tested in astronomy and (in general) in the laboratory
using bulk matter since  $\gamma_5$ couplings do not extend their influence
over
unpolarised macroscopic bo\-dies\footnote{This is true only to leading order in
the interaction as we shall show below.}. Of course, indirect limits in
macroscopic systems on the strength of such forces do exist, involving more
or less detailed knowledge on the physics of stellar evolution. The mass and
coupling strengths of axions for instance have been extensively surveyed and
constrained in various astrophysical or cosmological environments ranging
from bounds coming from cosmic string radiation of axions to bounds derived
from the cooling of white dwarfs and axion emission in Supernova collapse
\cite{e}.

In the present paper we would like to explore on a quantitative basis to what
extent are pseudoscalar long-range forces between unpolarised bodies
screened off over macroscopic scales. To this end we shall use astronomical
observations as well as laboratory experiments and we shall make no resort on
stellar
evolution modelling. We shall see, of course, as already mentioned above that
these
constraints are extremely weak  and even non existing in the astronomical
domain.

It is a well-known fact that spin dependent interactions do not
generally extend macroscopically over large distances. Indeed, one
would need polarised samples for forces to show their influence
coherently over macroscopic distances. We have in mind interactions such
as pseudoscalar particles coupled to nucleon sources (or, in general,
coupled to fermions) or Pauli magnetic moment couplings of fermions
to the electromagnetic field. In both cases, even though the range
of the interaction may be infinite (for a massless pseudoscalar and
obviously for the photon), macroscopic bulk matter does not feel those
forces unless the samples are polarised. Actually, it is the helicity-flip
nature of such interactions that prevents the exchanged quantum to extend
its influence
over macroscopic bodies. This fact, however, corresponds to the situation
where a single quantum is exchanged {``}at a time{''}, \ie it corresponds to
the ladder aproximation in a Bethe-Salpeter approach of the bound state
problem. Naturally then, the question arises as to what happens if two
quanta are exchanged {``}at a time{''}. That is, if we go beyond the ladder
approximation. Intuitively, the helicity non-flip nature is restored in
this case and, consequently, one should expect coherent effects to extend
over macroscopic matter. In this paper we explore this possibility and
find that indeed long-range macroscopic forces develop across extended
bodies. Of course, as one could also foresee, these effects are extremely
tiny.

We shall consider non-relativistic, non second quantised matter coupled
to a pseudo-scalar quantum field $\phi$ . To be
definite consider the interaction
\be
\mbox{\cal L}_{int}=\frac{g}{2M} \:
\chi^{\dagger} \, \vec{\sigma}  \, \chi \:\cdot \,\nabla \phi (\vec{r}\,)
\ee
which is the static non-relativistic (NR) limit of \mbox{$\overline{\psi}(x)
\gamma_{5}\psi (x) \phi (x)$} ($\chi $ are the Pauli spinors, $M$ is the
fermion
mass). Of course, if $\phi (x)$ were a scalar field then \mbox{$\overline{\psi}
(x) \psi (x)$} would be, in the NR-limit, just a number density of particles
and
hence bulk matter could act coherently as a source, whereas for the
spin-flip interaction in eq (1) this is not the case.\footnote{The
electromagnetic
analogue of eq (1) can be derived from the Pauli interaction $\frac{\mu}{2}
\:\overline{\psi}\sigma_{\mu  \nu}\psi\, F^{\mu \nu}$ where $\mu$ is the
magnetic
moment. It reduces to
 $\mbox{\cal L}_{int}=\mu\,\chi^{\dagger} \,
\vec{\sigma} \, \chi \:\cdot \,
\vec{B} $  in the NR-limit. }

Now, consider two point-like matter sources a distance R apart at positions
$\vec{R}_{1}$ and $\vec{R}_{2}$, $\vec{R}=\vec{R}_{1}-\vec{R}_{2}$. The
interaction hamiltonian is
\be
H_{int}=\int d^{3}\vec{r}\sum_{i=1,2}
\frac{g}{2M}\, \delta (\vec{r}-\vec{R}_{i}) \, \vec{\sigma}^{(i)} \cdot \!
\nabla \phi (\vec{r}\,)
\ee
where \mbox{$\phi (\vec{r}\,)=\int \frac{d^{3}
\vec{k}}{(2\pi)^{3}}\frac{1}{2\omega_{k}}[a_{k}e^{-i\vec{k} \cdot
\vec{r}}+a^{\dagger}_{k}e^{i\vec{k} \cdot \vec{r}}]$} and \mbox{$\omega_{k}
=\sqrt{\vec{k}^2+m^2}$} with $m$ the mass of the pseudo-scalar quantum.
Eventually we shall make $m\rightarrow 0$, since we are interested in
long-range effects.

In order to account for the mean effect of bulk matter we shall actually
calculate $T \! r(\rho H_{int})$ where $\rho$ is the density matrix operator
that describes the state of the system in spin space of the two-fermion
sources (we label the states that span the two-fermion space by $|i\rangle$,
$i=1,\ldots,4$).

To do the calculation we shall resort to {``}old fashioned{''} perturbation
theory.
Since the hamiltonian in eq.\/(2) creates (annihilates) $\phi$ quanta
from  (into) the vacuum, the expectation value of $H_{int}$ between
vacuum pseudo-scalar states vanishes. There are no linear effects in
$g$. But  second order perturbation can, in principle, connect the
vacuum with the  vacuum, for $H_{int}$ can create one pseudo-scalar
and, on second  application, annihilate it again into the vacuum.

A straightforward calculation then gives, to second order, for the interaction
energy between the two sources,
\be
\Delta E^{(2)}_{int}=\frac{g^{2}}{4M^{2}}
\:\vec{\sigma}^{(1)}\!\cdot \! \nabla \:\vec{\sigma}^{(2)}\!\cdot \! \nabla\:
\frac{e^{-mR}}{4\pi R}
\ee
where $\left(\vec{\sigma}^{(1)} \cdot\;\;\;\right)\left(\vec{\sigma}^{(2)}
\cdot\;\;\;\right)$
has to be understood as an operator that acts on the $2\times 2$
dimensional spin-space of the two-fermion system. We should point out that
to reach eq.\/(3) we have substracted off the (infinite) self-interaction
energies of each source, which furthermore are independent of $R$ and hence do
not contribute to the force.

Eq.\/(3) is a well-known result \cite{f} and, clearly, when we perform the
spin average
in the tensor product space of the two Pauli spinors that corresponds to
unpolarised sources, the interaction energy vanishes since
\be
Tr(\vec{\sigma}\otimes\vec{\sigma})=0
\ee
(the density matrix is just $
\frac{1}{4}I$). Thus, we expect no macroscopic effect to this order as
already advertised before.

We turn now to the next order in perturbation theory, \ie $O(g^{4})$
since only even powers of $H_{int}$ are permitted. This corresponds
to the emission and subsequent absorption of two quanta
\footnote{For work on long-range two-photon forces see \cite{g} and references
therein.}
. Again we encounter
in the calculation infinities associated to self-energies of the sources.
Indeed, there are, to this order, the self-energies arising from the
emission and absorption of one quantum by each source and emission and
absorption of two quanta by either source. We simply substract off
these infinities for they again do not depend on $R$ and lead to no force. But,
there are also divergences corresponding to the  case where one quantum is
exchanged between both sources and the other one  is emitted and reabsorbed by
source 1 or source 2, respectively. These  last terms, however, vanish when
performing the statistical average over  unpolarised samples.

The result, after statistical averaging is
\begin{eqnarray}
\overline{\Delta
E}^{(4)}_{int}&=&-\frac{2}{(2\pi)^{6}}\left(\frac{g}{2M}\right)^{4}\:
Tr\left\{\left[(\vec{\sigma} \cdot
\!\nabla \otimes \vec{\sigma} \cdot \! \nabla) \int\frac{d^{3}\vec{p}}
{\omega_{p}^{3}}\cos(\vec{p} \cdot \! \vec{R})\right]\right.\nonumber\\
& &\times\left.\left[(\vec{\sigma}\cdot \! \nabla
\otimes \vec{\sigma} \cdot\!\nabla)\int\frac{d^{3}\vec{p'}}{\omega_{p'}^{2}}
\cos(\vec{p'} \cdot \!\vec{R})\right]\right\}
\end{eqnarray}
which is certainly different from zero. Performing the integrals,
eq.\/(5) can be put in the form
\be
\overline{\Delta E}^{(4)}_{int}=-\frac
{1}{4\pi^{3}}\left(\frac{g}{2M}\right)^{4}\sum_{i,j=1}^{3}\left(\frac{\partial}
{\partial R_{i}}
\frac{\partial}{\partial R_{j}}K_{0}[mR]\right)\left(\frac{\partial}
{\partial R_{i}}\frac{\partial}{\partial R_{j}}\frac{e^{-mR}}{R}\right)
\ee
with $K_{0}$ the
modified Bessel function. In the limit $m\rightarrow 0$, we obtain
\be
\overline{\Delta E}^{(4)}_{int}=-\:\frac{g^{4}}{16\pi^{3}M^{4}}\:\frac{1}
{R^{5}}
\ee
for the interaction energy for two spin averaged fermion samples normalized to
one fermion per sample
\footnote{The corresponding formula for the electromagnetic
Pauli dipole interaction is $\overline{\Delta
E}^{(4)}_{int}=-\:\frac{\mu^{4}}{\pi^{3}}\:\frac{1} {R^{5}} $.}.
The resulting force between two pieces of matter is attractive, composition
dependent and very small.

We should note that the statistical averaging in eq.\/(5) involves
the introduction of a finite renormalization of the two-fermion wavefunction.
In fact, to the order we are working, the statistical weights $\omega_{i}$
in the density matrix $\rho$ are related to the lowest order density matrix
elements that enter in eq.\/(5) through a rescaling $\omega_{i}=
\omega^{(0)}Z_{i}^{-1}$, where $Z_{i}$ is the (finite) wavefunction
renormalization
of the perturbed states, \ie $Z_{i}^{-1}=\langle i|i\rangle$.

The new force should couple to baryons and/or leptons and hence the coupling be
proportional to an arbitrary combination of baryon and lepton quantum numbers.
Since
the force goes as $M^{-4}$, in order to maximize the effect, we shall in what
follows
couple the interaction fully to the electrons in matter and thus guarantee
that our
bounds are the most restrictive one can get.

Already from the $R$ dependence of the force one can suspect that, on
astronomical scales, the effect of pseudoscalar massless exchange will be
totally negligible. Indeed, a quantity that typically supplies constraints
on non-newtonian components of the force in the astronomical realm is the
perihelion of Mercury. In our case, the effect would be a perihelion shift of
the order $\Delta \phi \sim 2\times 10^{-54} {\alpha_g}^2
\frac{\mbox{\scriptsize \rm rad}}{\mbox{\scriptsize \rm
rev}}$ which gives no limit at all on $\alpha_g=\frac{g^2}{4\pi}$. This result
is completely general: no astronomical observation is able to put
limits on pseudoscalar long-range forces.

As to laboratory experiments, at present the best laboratory limits to
non-newtonian components of the force between macroscopic bodies come from the
null experiments of Hoskins et al. and Chen et al.  \cite{h}. In
particular, Hoskins et al. test deviations from the $\frac{1}{R^2}$ law
in the 2-5 cm range. These authors place a test mass inside a long and
hollow cylinder and monitor the torque on the mass as the distance
between the mass and the cylinder is varied. The mass in an infinitely
long cylinder would experience no gravitational force and hence one can
set direct limits on non gravitational effects by performing the
experiment. There are corrections, of course, due to end effects in a
finite cylinder, but these are small enough to allow detection  of a
force deviating from the $\frac{1}{R^2}$ law. The actual experiment
reached a sensitivity to acceleration of $2\times 10^{-11}
\frac{\mbox{\scriptsize \rm cm}}{\mbox{\scriptsize \rm s}^2}$
and used a 20 g test mass of high-purity copper and a 60
cm long cylinder of thickness 1 cm and an interior diameter of 6 cm
made of high-purity double-vacuum-melted stainless steel. No deviation
to the newtonian law was observed at the level of the sensitivity
stated above. We can use these results to try to constrain our
pseudoscalar force by calculating the radial acceleration towards the
wall of the tube caused by our putative pseudoscalar force on the test
mass suspended from the end of a torsion balance bar \be a=\frac{(\hbar
c)^5}{c^8}\frac{5}{\pi}\frac{\alpha_g^2}{ M_e^6} f_e^{\mbox{\scriptsize \rm
(sample)}}f_e^{\mbox{\scriptsize \rm (cyl)}}\rho^{\mbox{\scriptsize
\rm (cyl)}}
\; \int_{R_0}^{R_0+d} dr \,
\int_{0}^{2\pi} d\phi\,
\int_{-\frac{L}{2}}^{\frac{L}{2}} dz \,
\frac{r(r \cos{\phi}-R)}{(R^2+r^2+z^2-2 Rr \cos{\phi})^{7\over 2}}
\ee
where $M_e$ is the mass of the electron,
$f_e^{\mbox{\scriptsize \rm(sample)}}$
and $f_e^{\mbox{\scriptsize \rm
(cyl)}}$ are respectively the electronic mass fraction of the sample and the
one of the cylinder, $\rho^{\mbox{\scriptsize \rm (cyl)}}$
is the mass density of the cylinder,
$R_0$ is its interior radius, $d$ its wall thickness and $L$ its length, and
$R$
is the distance between the sample and the interior cylinder wall. Feeding the
actual values for these parameters and doing the integral, one can check that
the experimental results are compatible with an $\alpha_g < 1.27$. This bound
is very poor. Although it is many orders of magnitude better than the
astronomical bounds discussed before, again the conclusion is that these forces
are macroscopically very efficiently shielded, so that only detailed
microscopic
physics can provide useful information about them.

Finally, we would like to compare the effect of pseudoscalar forces with the
well known Casimir effect \cite{i}. This very tiny effect, due to
quantum fluctuations of the electromagnetic field, has been
experimentally verified  \cite{j} and we wish to quantify the strength
of our putative force compared to the Casimir force.

  Consider the force between
two large  parallel plates of thickness $d$ separated by a distance $a$.
Ignoring  edge effects (\ie $a$,$d$ much smaller than the size of the plates)
one  obtains for copper,
\be
\left(\frac{F}{S}\right)_{Cu}=-1.4\times
10^{-3} \, {\alpha_g}^2
\left(\frac{1}{(a+2d)^2}+\frac{1}{a^2}-\frac{2}{(a+d)^2}\right) \;\;N/m^2
\ee
for the attractive force per unit surface area where $a$,$d$ are in $\mu m$.
Of course, we took copper just to be definite but it could equally well be any
other material, conducting or dielectric.
Let us compare this to the Casimir force between the plates
(in this case the plates must be necessarily conducting)
\be
\frac{F_{Cas}}{S}=-1.3\times
10^{-3}\frac{1}{a^4}\;\;N/m^2
\ee
Notice that for very thin plates ($d\ll a$) the $a$ behaviour of both forces
is identical.

These results imply that a verification of the Casimir effect with, say, a 10\%
precision, demands $\alpha_g < 0.27$, \ie less than 1, but still a poor bound.

We note in passing that since $\alpha=\frac{1}{137}$ is always less than the
bounds for $\alpha_g$ obtained, the equivalent electron Pauli magnetic moment
effect is also undetectable.

To summarize, we have shown that residual long-range attractive
and composition dependent forces between macroscopic unpolarised
neutral bodies which are associated to spin dependent couplings of
fermions to massless bosons -a quantum mechanical effect due to the
exchange of two quanta {``}at a time{''}-  are very efficiently shielded
over an enormous range of distances: from astronomical scales down to
the micron scale.

\vspace{1cm}
\underline{Aknowledgement}

This work has been partially supported by CICYT under project number
\mbox{AEN-93-0474}.


\begin{thebibliography}{99}
\bibitem{a} For a review, see

E. Fischbach and C. Talmadge, Nature \underline{356} (1992) 207.

E. Fischbach, G.T. Gillies, D.E. Krause, J.G. Schwan and C. Talmadge,
Metrologia
\underline{29} (1992) 213.

\bibitem{b}J.Frieman and B. Gradwohl, Phys. Rev. Lett. \underline {67} (1991)
2926.

J. Frieman and B. Gradwohl, Ap. J. \underline {398} (1992) 407.

C.W. Stubbs, Phys. Rev. Lett. \underline {70} (1993) 119.

G. Smith, E.G. Adelberger, B.R. Heckel and Y. Su, Phys. Rev. Lett. \underline
{70} (1993) 123.

\bibitem{c}T.D. Lee and C.N. Yang, Phys. Rev.  \underline {98}  (1955) 50.

\bibitem{d}A.D. Dolgov and Ya.B. Zeldovich, Rev. Mod. Phys. \underline {53}
(1981)  1.

L.B. Okun, Yad. Fiz. \underline {10}  (1969)  358. [Sov. J. Nucl. Phys.
\underline {3}  (1966)  837].

J.E. Moody and F. Wilczek, Phys. Rev. \underline {D30}  (1984)  130.

A. de R{\'u}jula, Phys. Lett. \underline {180B}  (1986)  213.

C. Talmadge, J.P. Berthias, R.W. Hellings and E.M. Standish,  Phys. Rev. Lett.
\underline {61}  (1988) 1159.

\bibitem{e} For a review, see G.G. Raffelt, Phys. Rep. \underline {198}
(1990)  1.

\bibitem{f}J. J. Sakurai,
\mbox{\em Advanced Quantum Mechanics}, Addison-Wesley
(1967).

\bibitem{g} G. Feinberg and J. Sucher, Phys. Rev.
\underline{D45} (1992) 2493.

G. Feinberg, J. Sucher and C.K. Au, Phys. Rep. \underline {180} (1989)
83.


\bibitem{h} J.S. Hoskins, R.D. Newman, R. Spero and J. Schultz, Phys.
Rev. \underline {D32} (1985) 3084.

Y.T. Chen, A.H. Cook, A.J.F. Metherell, Proc. R. Soc. \underline {A394} (1984)
47.

\bibitem{i} H. B. G. Casimir, Proc. Kon. Ned. Akad. Wet.
\underline {51}  (1948) 793.

\bibitem{j} M.J. Sparnaay, Physica \underline {24} (1958) 751.

\end{thebibliography}
\end{document}